\documentstyle[12pt,fleqn]{article}

\input epsf

\def\singlespace {\smallskipamount=3.75pt plus1pt minus1pt
                  \medskipamount=7.5pt plus2pt minus2pt
                  \bigskipamount=15pt plus4pt minus4pt
                  \normalbaselineskip=15pt plus0pt minus0pt
                  \normallineskip=1pt
                  \normallineskiplimit=0pt
                  \jot=3.75pt
                  {\def\smallskip {\vskip\smallskipamount}}
                  {\def\medskip   {\vskip\medskipamount}}
                  {\def\bigskip   {\vskip\bigskipamount}}
                  {\setbox\strutbox=\hbox{\vrule 
                    height10.5pt depth4.5pt width 0pt}}
                  \parskip 7.5pt
                  \normalbaselines}
\def\middlespace {\smallskipamount=5.625pt plus1.5pt minus1.5pt
                  \medskipamount=11.25pt plus3pt minus3pt
                  \bigskipamount=22.5pt plus6pt minus6pt
                  \normalbaselineskip=22.5pt plus0pt minus0pt
                  \normallineskip=1pt
                  \normallineskiplimit=0pt
                  \jot=5.625pt
                  {\def\smallskip {\vskip\smallskipamount}}
                  {\def\medskip   {\vskip\medskipamount}}
                  {\def\bigskip   {\vskip\bigskipamount}}
                  {\setbox\strutbox=\hbox{\vrule 
                    height15.75pt depth6.75pt width 0pt}}
                  \parskip 11.25pt
                  \normalbaselines}
\def\doublespace {\smallskipamount=7.5pt plus2pt minus2pt
                  \medskipamount=15pt plus4pt minus4pt
                  \bigskipamount=30pt plus8pt minus8pt
                  \normalbaselineskip=30pt plus0pt minus0pt
                  \normallineskip=2pt
                  \normallineskiplimit=0pt
                  \jot=7.5pt
                  {\def\smallskip {\vskip\smallskipamount}}
                  {\def\medskip   {\vskip\medskipamount}}
                  {\def\bigskip   {\vskip\bigskipamount}}
                  {\setbox\strutbox=\hbox{\vrule 
                    height21.0pt depth9.0pt width 0pt}}
                  \parskip 15.0pt
                  \normalbaselines}

\begin{document}

\singlespace

\rightline{TIFR-TAP Preprint}

\begin{center}

{\Large {Gravitational Collapse and Cosmic Censorship\footnote{based 
on a talk given at the XVIIIth meeting of the
Indian Association for General Relativity and Gravitation,
Institute of Mathematical Sciences, Madras, India, 15-17 February, 1996;
to appear in the Conference Proceedings - Institute of Mathematical
Sciences Report  IMSc - 117, \ Editors: G. Date and\hfil\break B. R. Iyer}}} 

\end{center}
\vspace{1.0in}
\vspace{12pt}
\begin{center}
{\large{T. P. Singh\\
Theoretical Astrophysics Group\\
Tata Institute of Fundamental Research\\
Homi Bhabha Road, Bombay 400005, India.\\
tpsingh@tifrvax.tifr.res.in\\}}
\end{center}
\vskip 1 in
\centerline{\bf ABSTRACT}
\medskip

\noindent This article gives an elementary overview of the end-state
of gravitational collapse according to classical general relativity.
The focus of discussion is the formation of black holes and naked 
singularities in various physically reasonable models of gravitational
collapse. Possible implications for the cosmic censorship hypothesis
are outlined.

\vfil\eject

\centerline{\bf I. Introduction}

\medskip

\noindent It is expected that a very massive star will not end up
either as a white dwarf or as a neutron star, and that it
will undergo an intense gravitational collapse towards the end of its
life history. The very late stages in the evolution of such a 
star will necessarily be determined by quantum gravitational effects. Since we
do not yet have a quantum theory of gravity, we cannot give a definite
description of these very late stages. However, we can at least ask what
classical general relativity predicts for the final stages of the
evolution. It is possible that the answer given by the classical theory
will have some connection with the answer coming from quantum 
gravity - perhaps the latter will provide a sort of quantum correction
to the former.

It is remarkable that seventy years after Einstein proposed the general
theory of relativity we do not have a complete understanding of the
theory's prediction for the end-state of gravitational collapse. This
situation is intimately related with our lack of understanding of the
general global properties of the solutions of Einstein's equations. The
most significant developments to date in the study of gravitational collapse
have been the singularity theorems of Hawking and Penrose [1]. 
In the context of gravitational collapse,
the theorems show that if a trapped surface forms during the
collapse of a compact object made out of physically reasonable matter, the 
spacetime geometry will develop a gravitational singularity
(assuming the non-existence of closed time-like curves). By a 
gravitational singularity one means that
the evolution of geodesics in the spacetime will be incomplete. It is
plausible that the formation of a gravitational singularity in a collapsing
star will be accompanied by a curvature singularity - one or more
curvature scalars will diverge.

The general conditions which will ensure the formation of a trapped
surface are not well understood. This is one of the aspects in which our 
understanding of gravitational collapse is incomplete.
In this article, however, we will not be concerned with this particular 
issue. We will assume that a gravitational singularity does form, either 
because the conditions of the singularity theorems have been met, or 
otherwise. All the same it should be mentioned that the astrophysical
parameters for very massive collapsing stars are usually such that a trapped 
surface can be expected to form during gravitational collapse.

It maybe the case that the singularity is not visible to
a far-away observer because light is not able to escape the collapsing
star. This is essentially what we mean when we say that a black-hole
has formed. The singularity is hidden from view by the event horizon,
which is the boundary of that spacetime region surrounding the singularity
which cannot communicate with the far-away observer (Figure 1). 

The singularity theorems of Hawking and Penrose do not imply that the
collapsing star which develops a singularity will necessarily become a 
black-hole. That is, even if a trapped surface and hence a singularity does
form during collapse, the trapped surface need not hide the singularity from
a far-away observer. This alternate possibility is called a naked singularity
(Figure 2). In this case, the event-horizon fails to cover the 
singularity and light from the singularity could escape to infinity. One 
could form a rough picture of this situation by imagining that a singularity
has formed at the center of a collapsing spherical star {\it before} its 
boundary has entered the Schwarzschild radius. Such a singularity could 
be visible to an observer watching the collapse. Also, if a
singularity forms without the conditions of the singularity theorems having
been met, it could be naked, as the event horizon may not form at all.

\begin{center}
\leavevmode\epsfysize=4 in\epsfbox{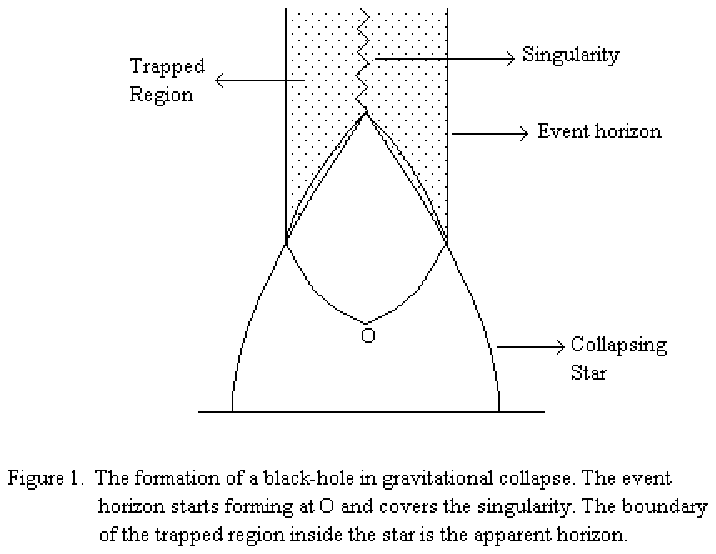}
\end{center}

Does gravitational collapse end in a black-hole or a naked singularity?
Einstein's equations must give a definite answer to this question, but
at present we do not know what that answer is. One could well ask, what 
difference does it make? The difference is very significant. If a 
singularity is naked, one
could prescribe arbitrary data on the singular surface - this would result in
total loss of predictability in the future of the singularity. If a
singularity is hidden behind an event horizon, predictability would be
preserved at least in the spacetime region outside the horizon. 

Furthermore, the issue has great importance for 
black-hole astrophysics and for the theory of black-holes. 
In terms of their astrophysical properties, naked singularities could
be very different from black-holes. While (classical) black-holes are
one-way membranes and inflowing matter simply gets sucked into the
singularity, quite the opposite may hold for a naked singularity; matter
might be thrown out with great intensity from near a naked singularity.
The validity of many theorems on black-hole
dynamics depends on the assumption of absence of naked singularities. 
The extensive and successful applications of black-holes in astrophysics,
and the detailed studies of their profound and elegant properties does lend
strong support to the belief in their existence. Nonetheless, it is an open
question as to whether gravitational collapse necessarily ends in a 
black-hole or could in some cases lead to a naked singularity.
If the latter is the case, then one needs to know what kind of stars end
as naked singularities.

\begin{center}
\leavevmode\epsfysize=4 in\epsfbox{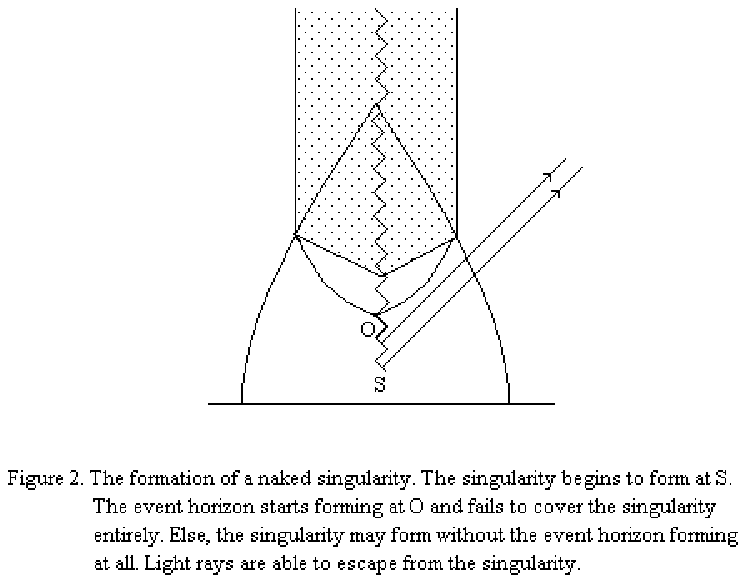}
\end{center}

In the absence of an answer to the above question, the basis for 
black-hole physics and its applications is the Cosmic Censorship Hypothesis.
In somewhat non-rigorous terms, the hypothesis could be stated as

\begin{itemize}
\item Gravitational collapse of {\it physically reasonable matter} 
starting from {\it generic}\hfil\break {\it initial data} leads to the
formation of a black-hole, not a naked singularity.
\end{itemize}

\noindent We will elaborate in a moment on the italics. Such a 
hypothesis was first considered by Penrose in [2], where he asked
as to whether there exists a Cosmic Censor who would always cloth a 
singularity with an event horizon. (It is interesting, though probably not
well-known, that on occasion Penrose has also written in support of naked
singularities [3]).
The censorship hypothesis has remained unproved despite many serious
efforts (for a review of some of the attempts see [4]). Part of the 
difficulty lies 
in not having a unique rigorous statement one could try to prove.
Studying gravitational collapse using Einstein equations is a formidable 
task, and evidence for {\it formation}
of a black-hole is as limited as for a naked singularity. This is simply
because very few examples of dynamical collapse have been worked out,
and not many exact solutions are known.
Until recently, the widely held belief has been that the hypothesis must be
right; however studies over the last few years have caused at least some
relativists to reconsider their earlier view. It is obvious, though not
often emphasized, that falsification of the hypothesis does not necessarily
rule out black-holes altogether but might allow only a subset of the
initial data to result in naked singularities. The former alternative
(i.e. no black-holes) is extremely unlikely, given the great richness of 
black-hole physics and astrophysics.

By `physically reasonable matter' one usually means that the (classical) 
matter obeys one or more of the energy conditions (for a discussion see [5]).
The weak energy condition for instance requires that the pressures in
the collapsing matter not be too negative, and the energy density be 
positive. By `generic initial data' one means that the solution of Einstein
equations being used to study collapse has as many free functions as are
required for the initial data to be arbitrary.

Many people consider naked singularities to be a disaster for general
relativity and for physics as such. We have mentioned that naked
singularities could result in a total loss of predictability to their
future. However, if we make the natural assumption that the ultimate theory
of gravity will preserve predictability (in some suitable sense of the 
concept), the
occurence of naked singularities in general relativity will signal a real
need for modification of the theory. Such a definite (and rare) signal
could only help improve our understanding of gravitation, instead of
being a disaster! To many other people, naked singularities are a positive
asset for astrophysics - the possibility of emission of light from high
curvature regions close to the singularity might make such singularities
extraordinary energy sources. Sometimes a view is expressed that in any
case a quantum theory of gravity will avoid singularities altogether - in
that case how does it matter whether the classical theory 
(general relativity) predicts the singularities to be naked or covered?
Our discussion above suggests that even if quantum gravity were to avoid
singularities, the spacetime regions that are naked according to the
classical theory will behave very differently from black-holes in the
quantized theory as well. Besides, the relevance of the hypothesis to 
black-hole astrophysics cannot be overlooked.

Perhaps it is important to mention that the validity of the hypothesis
could be discussed at two distinct levels. Firstly, we need to find out
if it holds in general relativity. However, even if relativity theory
allows naked singularities, it could be that actual stars may not end up as
naked singularities. This could happen if the initial conditions necessary
for a naked singularity to form are simply not observed in the real world.
Thus violation of the hypothesis at a theoretical level might compel us
to consider replacing general relativity by a better theory, but may not
have observational consequences.

The failure to prove the hypothesis has led, over the last ten years or
so, to a change in the approach towards the problem. Attention has
shifted to studies of specific examples of gravitational collapse. For
instance, people have been studying spherical gravitational collapse,
with a particular choice of the matter stress tensor - like dust, perfect
fluids, and massless scalar fields. These and other examples often show
that collapse of physically reasonable matter can end either in a 
black-hole or a naked singularity, depending on the choice of initial data.
To a degree, such examples go against expectations that the 
censorship hypothesis is correct. It, however, does remain to be seen as to 
whether or not generic initial data will lead to naked singularities.
These examples can be regarded as good learning exercises - at the very 
least we learn about the properties of naked singularities which form.
Perhaps these very properties might suggest that the singularity, though
naked, does not violate the spirit of the hypothesis.

In this article we will attempt an overview of 
the end state of gravitational collapse according to general relativity,
the focus of discussion being the censorship hypothesis. Rather than 
reviewing the attempts to prove the hypothesis, we will concern ourselves
with recent studies of models of collapse. It is relevant to note that
these models typically examine the properties of curvature singularities
that form during collapse. If a naked curvature singularity does form,
the issue of its geodesic incompleteness is to be handled separately - an
aspect we will consider briefly towards the end of the review. Also, we will
not discuss examples of static naked singularities in general relativity,
although many such are known.

In Section II, spherical gravitational collapse for various forms of 
matter is reviewed.
We also discuss some properties of the naked singularities found in these
models. Section III is a brief discussion of the limited results on 
non-spherical collapse. In the last section, a critical comparison of the
various existing results and their interpretation is attempted.

\vskip 0.2 in

\centerline{\bf 2. Spherical Gravitational Collapse}

\smallskip
\noindent It is only natural that most of the examples studied are for 
the idealized case of spherical collapse, in asymptotically flat backgrounds.
Even here, one does not yet know the conditions for formation of 
black-holes and naked singularities. Some results are known for specific 
forms of energy-momentum tensors. Often there are striking similarities
amongst results for different kinds of matter, suggesting an
underlying pattern. We review results for collapse of dust, perfect fluids
with pressure, a radiating star described by the Vaidya metric, massless
scalar fields, and spherical collapse for matter with no restriction
on the energy-momentum tensor, except the weak energy condition. 

\smallskip

\noindent{\bf 2.1 Dust collapse}

\noindent By dust one means a perfect fluid for which the pressure is
negligible and is set to zero. This highly idealized description has
the advantage that an exact solution of Einstein equations is known,
which describes the collapse of a spherical dust cloud in an asymptotically
flat spacetime. This is the Tolman-Bondi solution, given independently by
the two authors [6]. The evolution of the cloud is determined once the
initial density and velocity distribution of the fluid has been given.
Because of spherical symmetry these distributions will be functions only
of the radial coordinate $r$, so that the initial data consists of two
arbitrary functions of $r$. The cloud is assumed to extend upto a 
finite radius and the interior dust solution is matched to
a Schwarzschild exterior. Since we are interested in collapse, the
velocity of each fluid element is taken to be towards the center of the cloud.
It can be shown that starting from regular initial data, the collapse
leads to the formation of a curvature singularity.

A very special case of the Tolman-Bondi solution is a dust cloud whose
initial density distribution is homogeneous, and the velocity increases
linearly with the physical distance from the center. This of course is
the Friedmann solution matched to a Schwarzschild exterior and was used
by Oppenheimer and Snyder [7] to provide the first example of dynamical 
collapse in general relativity. Starting from regular initial data, the
cloud develops a curvature singularity at its center which is not visible.
This was the first theoretical example of black-hole formation (Figure 3(i)).

For many years, the work of Oppenheimer and Snyder has remained a model
for how a black-hole might form in gravitational collapse. The collapsing
star will enter its Schwarzschild radius, become trapped and proceed to
become singular, and the singularity will be hidden behind the event
horizon. The censorship hypothesis was also originally inspired essentially
by this work, because not much more was known anyway about properties
of gravitational collapse. In hindsight, one might be surprised with the
degree of generality attributed to results arising from the
study of a model star that is spherically symmetric, homogeneous and made 
of dust - none of the three properties are obviously true for a real star!

\begin{center}
\leavevmode\epsfysize=4 in\epsfbox{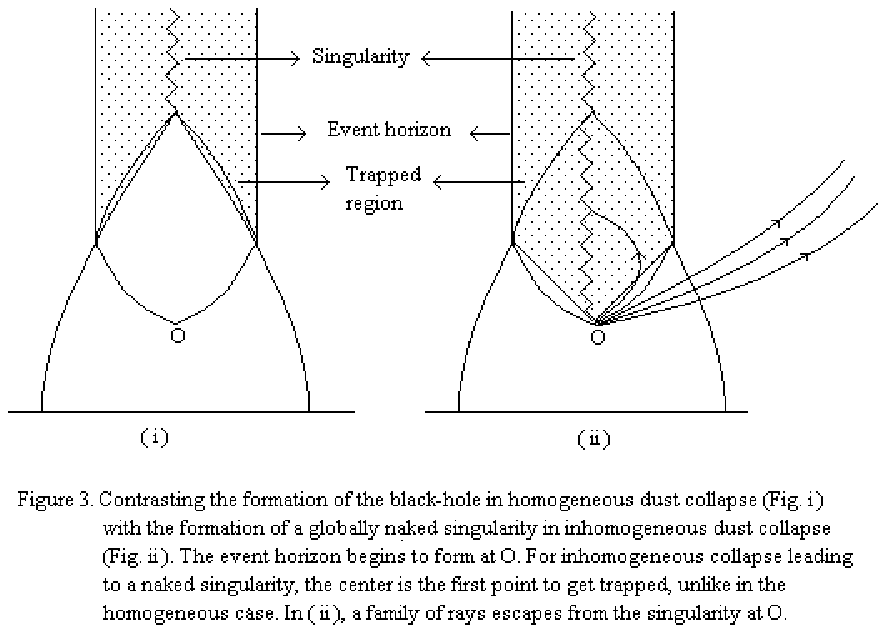}
\end{center}

Since the exact solution of Tolman and Bondi was known, it would have
been quite natural to extend the work of Oppenheimer and Snyder to
{\it inhomogeneous} dust collapse, described by this exact solution. However,
for nearly three decades after the paper of Oppenheimer and Snyder
was published, very
little work appears to have been done on gravitational collapse. Presumably,
there were not enough reasons for interest in the subject until the
discovery of quasars in the sixties, and the development of singularity
theorems. It is also true though that analysis of light propagation in 
the collapsing inhomogeneous dust star is a difficult task, and tractable
methods have been developed only in recent years.
The first work to deal with the inhomogeneous Tolman-Bondi
model appeared in the seventies [8]. Since then, many investigations of
inhomogeneous dust collapse have taken place [9], from the point of view
of the censorship hypothesis, and the last word on this particular model
has not yet been said. 

Yodzis et al. [8] were investigating what are called shell-crossing
singularities (or caustics) which form due to the intersection of two 
collapsing dust-shells at a point other than the center. These are curvature
singularities and they are also naked - but are not regarded as
violation of censorship since there is evidence [10] that such singularities
are gravitationally weak (as discussed later in the article). 
They are similar to the shell-crossing singularities that occur in
Newtonian gravity also and it is believed that spacetime can
be extended through such singularities. 

Of a more serious nature are the
shell-focussing singularities which form at the center of the cloud - they
result from the shrinking of collapsing shells to zero radius.
It was found by various people that for some of the initial density and
velocity distributions, the collapse ends in a naked singularity, whereas
for other distributions it ends in a black-hole. Also, 
both the black-hole and naked singularity solutions result from a non-zero
measure set of initial data. In particular, there was found a one-parameter 
family of solutions (described say by the parameter $\xi$), such that for 
$\xi < \xi_{c}$ the collapse leads to a black-hole, whereas for
$\xi > \xi_{c}$ it leads to a visible singularity.

The space-time diagram for inhomogeneous dust collapse leading to a naked
shell-focussing singularity is shown in Figure 3(ii), and should be 
contrasted with 
Figure 3(i) for Oppenheimer-Snyder collapse. Of particular importance is the
difference in the evolution of trapped surfaces in the two cases, and the
fact that for inhomogeneous collapse different shells become singular
at different times, unlike in the homogeneous case. For a discussion of
trapped surfaces in dust collapse see Jhingan et al. in [9]. 

At this stage we need to distinguish between a {\it locally} naked
singularity and a {\it globally} naked singularity. We say the singularity
is locally naked if light-rays do emerge from the singularity but fall back
to the center without escaping the event-horizon. Such a singularity will be
visible to an infalling observer who has entered the Schwarzschild radius,
but cannot be seen by an asymptotic observer (Figure 4). A singularity
is called globally naked if light-rays emerging from the singularity escape
the event-horizon and reach an asymptotic observer. We say
the singularity is visible if there are light-rays emerging from the 
singularity - a visible singularity may be locally or globally naked.
We say that a piece of the singularity is covered if it is not even locally
naked. In our terminology, a locally naked singularity is also a black-hole.
In this article when we call a singularity naked, we mean it is globally 
naked. The {\it weak} censorship hypothesis allows for the occurence of 
locally naked singularities but not globally naked ones, whereas {\it strong} 
censorship does not allow either. 

\begin{center}
\leavevmode\epsfysize=4 in\epsfbox{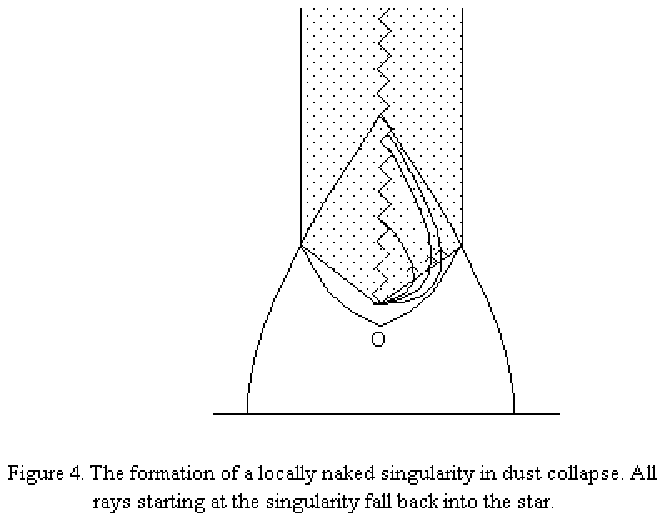}
\end{center}

As an illustration we describe in some detail the gravitational collapse of
an inhomogeneous dust cloud, starting from rest. It can be shown that the
singularity resulting from the collapse of shells with $r>0$ is covered.
At most, the singularity forming at $r=0$ (the central singularity) can be
naked. The conditions for the central singularity to be naked are given as
follows. Let the initial density $\rho(R)$ as a function of the physical
radius $R$ be given as a power-series, near the center:
$$ \rho(R) = \rho_{0} + {1\over 2}\rho_{2} R^{2} +
             {1\over 6}\rho_{3} R^{3} +   ...$$
where $\rho_{0}$ is the initial central density, and $\rho_{2}$ and 
$\rho_{3}$ are respectively the second and third derivatives at the center.
We assume that the density decreases with increasing $R$, hence the first
non-vanishing derivative is negative. It turns out that if $\rho_{2}<0$
the singularity is visible. If $\rho_{2}=0$ and $\rho_{3}<0$
then one defines a parameter $\xi=|\rho_{3}|/\rho_{0}^{5/2}$. The singularity
is visible for $\xi>25.47$ and covered for $\xi<25.47$. If 
$\rho_{2}=\rho_{3}=0$ the singularity is covered and we
have the formation of a black-hole. The Oppenheimer-Snyder example is a
subset of this case. In the case of a visible singularity,
entire families of light-rays emerge from the singularity.
Note that $\rho_{2}<0$ is generic and $\rho_{2}=0$
non-generic, hence dust collapse starting from rest leads to a visible 
singularity for generic initial density profiles. 

We find that there is a transition from naked
singularity type behaviour to black-hole type behaviour as the density
profile is made less inhomogeneous by setting more and more density
derivatives to zero. In the cases where the singularity is visible, the 
initial density distribution through the star then determines whether 
the singularity is locally or globally naked - examples of both kind occur.
When the collapsing dust cloud has an initial velocity, the overall picture
regarding the nature of the singularity is essentially similar to the
one described here for the case of a cloud collapsing from rest. 
Both black-holes and visible singularities now arise from generic initial
data. A Penrose diagram for the naked singularity is shown in Figure 5. In 
summary, when one allows for inhomogeneity in the density distribution, the 
nature of gravitational collapse is quite different from what Oppenheimer and 
Snyder found for the homogeneous case. Some open issues relating to dust 
collapse are pointed out later, in the discussion.

\begin{center}
\leavevmode\epsfysize=4 in\epsfbox{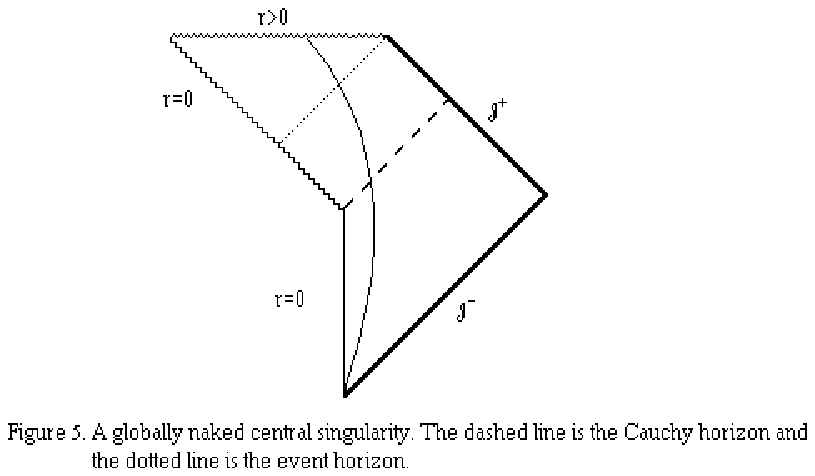}
\end{center}

\smallskip

\noindent{\bf 2.2 Including Pressure}

\smallskip

\noindent The description of collapsing matter as dust is an idealization and
a realistic study of collapse must take pressure gradients into account.
However, useful exact solutions of Einstein equations describing spherical
collapse of matter with pressure are hard to come by, and as a result a
clear picture of the kind presented above for dust does not yet exist. It
is in fact remarkable that we do not even know how the Tolman-Bondi solution
would change under the introduction of ``small'' pressures. For instance
if we are considering an equation of state $p=k\rho$ for a perfect fluid,
with $k$ very close to zero, it is not clear whether the solution is a
perturbation to the dust solution. In this section we summarize the few known
results on collapse of relativistic fluids, vis a vis the censorship
hypothesis.

When pressure is included, the energy momentum tensor $T_{ik}$ for matter 
undergoing spherical collapse is conveniently described in a comoving 
coordinate system. In these coordinates, $T_{ik}$ is diagonal and its
components are the energy density, the radial pressure and the tangential
pressure. (The only exception to this general description of the matter is the
case of null dust, described by the Vaidya spacetime, and reviewed
below). For a perfect fluid, the two pressures are identical. Since
perfect fluids are easier to study compared to imperfect ones, they have
inevitably received greater attention. 

An important early paper on the collapse of perfect fluids is that of 
Misner and Sharp [11]. They set up the Einstein equations for this system
in a useful physical form, bringing out the departures from the 
Oppenheimer-Volkoff equations of hydrostatic equilibrium. However, 
they did not consider solutions of these equations. Lifshitz and
Khalatnikov, and Podurets [12] worked out the form of the solution 
near the singularity for the equation of state of radiation, $p=\rho/3$.
Their approach has not received much attention, but appears to offer
a promising starting point for investigating censorship.

A significant development was the work of Ori and Piran [13], who 
investigated the self-similar gravitational collapse of a perfect fluid
with an equation of state $p=k\rho$. It is readily shown that the collapse
leads to the formation of a curvature singularity. The assumption of 
self-similarity reduces Einstein equations to ordinary differential 
equations which are solved numerically, along with the equations 
for radial and non-radial null geodesics. It is then shown that for every 
value of $k$ (in the range investigated: $0\leq k\leq 0.4$) there are
solutions with a naked singularity, as well as black-hole solutions.
Each kind of solution has a non-zero
measure in the space of spherical self-similar solutions for this 
equation of state. The issue as to which initial data (density profile,
velocity profile, and value of $k$) leads to naked singularities has not
been fully worked out though, and is an interesting open problem. (We have
in mind a comparison with the results for dust, quoted above).

An analytical treatment for this problem was developed by Joshi and
Dwivedi [14]. After deriving the Einstein equations for the collapsing
self-similar perfect fluid they reduce the geodesic equation, in the
neighborhood of the singularity, to an algebraic equation. The free
parameters in this algebraic equation are in principle determined by
the initial data. The singularity will be naked for those values of the
parameters for which this equation admits positive real roots. Since
this is an algebraic equation, it will necessarily have positive roots
for some of the values of the parameters, and for the initial data
corresponding to such values of the parameters the singularity is
naked. It is shown that families of non-spacelike geodesics will emerge
from the naked singularity. As in the case of Ori and Piran's work, the 
explicit relation between the initial data and the naked singularity
has not yet been worked out. Also it is not clear as to what is the 
measure of the subset of solutions leading to a naked singularity or a
black-hole. This analysis was extended to a self-similar spacetime with
a general form for $T_{ik}$ [15].

On physical grounds, imperfect fluids are more realistic than perfect
ones; very little is known about their collapse properties though. An
interesting paper is the one by Szekeres and Iyer [16], who do not 
start by assuming an equation of state. Instead they assume the metric
components to have a certain power-law form, and also assume that collapse
of physically reasonable fluids can be described by such metrics. The
singularities resulting in the evolution are analysed, with attention
being concentrated on shell-focussing singularities at $r>0$. They find that
although naked singularities can occur, this requires that the radial or
tangential pressure must either be negative or equal in magnitude to the
density. 

Lastly we mention the collapse of null dust (directed
radiation), described by the Vaidya spacetime [17]. Since the exact solution
is known, the collapse has been thoroughly investigated [18] for the
occurence of naked singularities. One considers an infalling spherical 
shell of radiation and imagines it as being made of layers of thin shells.
A thin shell becomes singular when its radius shrinks to zero. Let the
shells be labelled by the advanced time coordinate $v$, with $v=0$ for
the innermost shell, and let $m(v)$ be the Vaidya mass function. It can be
shown that the singularity at $v>0$ is covered. For $v=0$ the singularity
is naked if $\lambda\equiv 2\; dm(v)/dv|_{v=0}$ is less than or equal
to $1/8$, and covered otherwise. These results bear an interesting
similarity with those for dust, described in the previous section.

It will be evident from the previous few paragraphs that our understanding
of spherical collapse, when pressure gradients are included, is rather
incomplete. Ultimately, one would like to develop the kind of clear picture
that is available for dust collapse. It is interesting however that the
collapse of a self-similar perfect fluid, and of the fluids
considered by Szekeres and Iyer, admits both black-hole and naked
singularity solutions. This also brings forth an astrophysical 
issue - what is the relevant equation of state in the final stages of
collapse of a star? Could it be that the initial data leading to a naked
singularity is not being realised astrophysically? 

\bigskip

\noindent{\bf 2.3 Collapse of a scalar field}

\smallskip

\noindent One could take the viewpoint that the description of matter as a
relativistic
fluid is phenomenological, and that the censorship hypothesis should be 
addressed by considering matter as fundamental fields. As a first step, one
could study the spherical collapse of a self-gravitating massless scalar
field. A good deal of work has been done on this problem over the last
few years, and exciting results have been found. Here we can provide only
a very brief overview.

In a series of papers, Christodoulou has pioneered analytical studies of
scalar collapse [19]. He established the global existence and uniqueness
of solutions for the collapsing field, and also gave sufficient conditions
for the formation of a trapped surface. For a self-similar scalar collapse
he showed that there are initial conditions which result in the formation
of naked singularities. He has claimed that such initial conditions are
a subset of measure zero and hence that naked singularity formation is an
unstable phenomenon.

Christodoulou was also interested in the question of the mass of the
black-hole which might form during the collapse of the scalar wave-packet.
Given a one parameter family ${\cal S}[p]$ of solutions labelled by the 
parameter $p$
which controls the strength of interaction, it was expected that as $p$
is varied, there would be solutions with $p\rightarrow p_{weak}$ in which
the collapsing wave-packet disperses again, and solutions with
$p\rightarrow p_{strong}$ which have black-hole formation. For a given
family there was expected to be a critical value $p=p_{*}$ for which the
first black-hole appears as $p$ varies from the weak to the strong range.
Do the smallest mass black holes have finite or infinitesimal mass [20]?
This issue would be of interest for censorship, since an infinitesimal
mass would mean one could probe arbitrarily close to the singularity.
Of course when one is considering real stars, a finite lower limit on the
mass of the collapsing object arises because non-gravitational forces are
also involved. 

This problem was studied by Choptuik [21] numerically and some remarkable
results were found. He confirmed that the family ${\cal S}[p]$
has dispersive solutions as well as those forming black-holes, and a 
transition takes place from one class to the other at a critical $p=p_{*}$.
The smallest black-holes have infinitesimal mass. Near the critical
region, the mass $M_{bh}$ of the black-hole scales as 
$M_{bh}\approx (p-p_{*})^{\gamma}$, where $\gamma$ is a universal constant
(i.e. same for all families) having a value of about 0.37. The near critical
evolution can be described by a universal solution of the field equations
which also has a periodicity property called echoing, or discrete 
self-similarity. That is, it remains unchanged under a rescaling
$(r,t)\rightarrow (e^{-n\Delta}r, e^{-n\Delta}t)$ of spacetime coordinates.
$n$ is an integer, and $\Delta$ is about 3.4. Subsequently, these results
have been confirmed by others [22]. The occurence of black-holes with
infinitesimal mass goes against the spirit of censorship. The critical
solution $(p=p_{*})$ is a naked singularity. However, since the naked
singularity is being realised for one specific solution in the one parameter
family, it is a measure zero subset.   

Attempts have been made to construct analytical models which will
reproduce Choptuik's numerical results [23]. Since it is difficult
to make a model with discrete self-similarity, continuous self-similarity
is assumed instead. Brady showed that there are solutions which have
dispersal, as well as solutions which contain a black-hole or a naked
singularity. It would be of interest to relate his results to the naked
singularity solutions found by Christodoulou for self-similar collapse.
Recently, Gundlach has constructed a solution with discrete self-similarity,
which agrees with the critical universal solution found numerically by
Choptuik [24].

Similar critical behaviour has also been found in numerical studies
of collapse with other forms of matter. Axisymmetric collapse of
gravitational waves was shown to have a $\gamma$ of about 0.36, and
$\Delta \simeq 0.6$ [25]. For spherical collapse of radiation (perfect
fluid with equation of state $p=\rho/3$) the critical solution has
continuous self-similarity, and $\gamma$ of about 0.36 [26]. However it
has become clear now that the critical exponent $\gamma$ is not independent
of the choice of matter.
A study of collapse for a perfect fluid with an equation of state 
$p=k\rho$ shows that $\gamma$ depends on $k$ [27]. For a given form of
matter, there appears to be a unique $\gamma$, but the value changes as
the form of $T_{ik}$ is changed. Further studies of critical behaviour
are reported in [28]. 

The models described in this section exhibit a naked-singularity like
behaviour for near-critical solutions - such solutions are presumably
of measure zero on the space of all solutions. In the supercritical 
region the collapse is said to lead to the formation of a black-hole.
This raises a question as to how these results, say the supercritical
solutions for radiation fluid collapse, are consistent with those of Ori and 
Piran, who do find naked singularities. (The Ori-Piran naked singularity lies
in the supercritical region). It maybe that when one numerically finds
a singularity at the center $r=0$ one is not easily able to tell whether
this is a black-hole or a naked singularity, and this may have to be
investigated further. 

It is perhaps also relevant to note that collapse of real stars which
proceed to become singular is expected to be described by supercritical 
solutions. Thus the naked singularity observed near the critical region,
while of major theoretical interest, may not have astrophysical 
implications. This further emphasizes the need for investigating whether
the singularity being observed numerically in the supercritical region
is necessarily covered by the horizon, or could be naked.

It is undoubtedly true that these studies of critical behaviour have
opened up an entirely new aspect of gravitational collapse and the
related issue of censorship. Obtaining a theoretical understanding of
these numerically observed phenomena is an important open problem.

\bigskip

\noindent{\bf 2.4 Spherical collapse with general form of matter}

\smallskip

One finds a certain degree of similarity in the collapse behaviour of
dust, perfect fluids and scalar fields - in all cases some of the initial
distributions lead to black holes, and other distributions lead to
naked singularities. This would suggest an underlying pattern which is
probably characterized, not by the form of matter, but by some invariants
of the gravitational field. The role of matter may simply be that of 
determining which part of the parameter space these invariants lie in. 
Hence studies of collapse which put no restriction on $T_{ik}$ apart from
an energy condition should prove useful (still maintaining spherical
symmetry).

An interesting attempt in this direction has been made by
Dwivedi and Joshi in [29], where they generalized their earlier work
on dust collapse and self-similar fluids. They assumed a general $T_{ik}$
obeying the weak energy condition, and also that the collapsing matter
forms a curvature singularity at $r=0$ (the central singularity).
As we noted earlier, in the comoving coordinate system, matter is 
described by its energy density and the radial and tangential 
pressures. Along with these three functions, three functions describing
the metric enter a set of five Einstein equations, which are augmented
with an equation of state in order to close the system. The geodesic
equation for radial null geodesics is written in the limit of approach
to the singularity, and it is shown that the occurence of a visible
singularity is equivalent to the occurence of a positive real root for
the geodesic equation, suitably written. Since this equation depends on
free initial data, it follows that for a subset of the initial data there
will be positive real roots and the singularity will be visible.

This approach needs to be pursued further in order to find out whether
or not the naked singularities are generic. Also, it is of interest to
work out as to exactly which kind of initial data lead to naked 
singularities. These are difficult problems, in the absence of known
exact solutions. Another interesting attempt at treating spherical
collapse without restricting $T_{ik}$ is due to Lake [30]. He concluded
that a visible central singularity could form if the mass function
in the neighborhood of the singularity satisfies certain conditions.
The relation of these conditions with the initial data is not yet
apparent.

\bigskip

\noindent{\bf 2.5 Properties of naked singularities}

\smallskip

\noindent It is evident that energy conditions by themselves do not
restrict the occurence of naked singularities. One would then like to
examine in some detail properties of these naked singularities, so as
to see if these properties might contain clues for a censorship 
hypothesis. We review below some of the important features of the naked
singularities found in various models.

\noindent{\it Curvature strength}:
When a collapsing star develops a curvature singularity, the 
energy density
also becomes singular. However, finite physical volumes may or may not be 
crushed to zero volume as the singularity is approached. This could be
used as a criterion for judging the physical seriousness of the singularity,
and also for the possible extendibility of spacetime through
the singularity [31]. We call a singularity a {\it strong}
curvature singularity if collapsing volume elements do get crushed to zero
at the singularity, and a {\it weak} curvature singularity if they do not.
(The terms {\it weak} and {\it strong} singularity are sometimes used in the
literature with a different meaning. We use them here in Tipler's sense).
It is believed that spacetime cannot be extended
through a strong singularity, but is possibly extendible through a weak   
one. A rigorous proof for this is not yet available but is being attempted
by some relativists (for detailed studies see [32]). Clarke and Krolak [33]
gave necessary and sufficient quantitative criteria for the singularity to be 
strong, in terms of the rate of growth of curvature along outgoing 
geodesics, as the singularity is approached. 
 
A strong naked singularity is regarded as a more
serious violation of censorship as compared to a weak one. For instance, 
the shell-crossing type singularities are gravitationally weak [10].
Newman [34] studied the naked central singularity in Tolman-Bondi dust
collapse for a wide class of initial data and showed it to be weak. On this
basis it was conjectured in [34] that strong naked singularities do not
occur in collapse. It was however shown by various people [35] that
inclusion of initial data not considered by Newman gives rise also
to strong central naked singularities in dust. Interestingly, it has
recently become clear that the initial data leading to these strong naked
singularities is non-generic, whereas the data leading to a weak naked
singularity is generic [36]. In this sense Newman's conjecture does hold
for dust collapse! But strong naked singularities have been found in
other models - for instance in the naked singularities in the Vaidya
spacetime, where they arise from generic initial data. They were also
found by Ori and Piran in their study of the self-similar perfect fluid.
No results on strength seem to be known for scalar collapse.
The general $T_{ik}$ studied by Dwivedi and Joshi would lead to a strong
naked singularity for some initial data - however the genericity of such
initial data is an open issue. Thus the generality of strong
naked singularities remains unclear and it still might be possible
to formulate a censorship hypothesis along the lines of Newman's conjecture.

\noindent{\it Are naked singularities massless?}:
In all known examples of naked singularities, the mass of the collapsing
object (well-defined in spherical symmetry, with a vacuum exterior) is found
to be zero at the point where the singularity forms. There is evidence
that this is a general property of naked singularities in spherical
collapse [30]. On the other hand the black hole singularity is always found
to be massive. Since a massless singularity might be thought of as having
no associated gravitational field, this has led to the suggestion that such
singularities do not violate censorship. Note however that even from this
{\it massless} naked singularity entire families of geodesics emerge,
and it is not clear whether it is the mass or the outgoing geodesics which
are a more important property of the naked singularity!

\noindent{\it Redshift}: In known examples of naked singularities for
dust and perfect fluids, the redshift along outgoing geodesics 
emerging from the singularity is found to be infinite (when calculated
for observers in the vacuum region). This could be interpreted to mean
that no ``information'' is being transmitted from the naked singularity
and could be yet another approach to preserving censorship.

\noindent{\it Stability and Genericity}: This of course is the most
important issue relating to the naked singularity examples, and a notion
of stability is hard to define. The most direct definition of
stability (equivalently, genericity) of naked singularities would be
simply as follows. If a solution of Einstein equations describing collapse
leading to a naked singularity has as many free functions as required for
arbitrary initial data, the solution is stable. (One is reminded here of the
methods adopted by Belinskii, Lifshitz and Khalatnikov to show that general
solutions of Einstein equations contain singularities). Of course progress
in such a broad sense is hopelessly difficult, and one talks of stability
of a given solution under specific kinds of perturbations. For instance,
one would consider stability of the solution against change of initial data,
against change of equation of state, and against non-spherical metric
perturbations. From these viewpoints, very little is known about the
stability of the naked singularity models mentioned in this article.
(It is important to note that equally little is known about the stability
of the black-holes which form in these models during collapse!
Various studies show that the event horizon is stable to small
perturbations [37], and hence the singularity is stable, but it could be 
either naked or covered.)

One very useful way to address stability of naked singularities is to
study the blue-shift instability of the Cauchy horizon. One considers
ingoing waves starting from null infinity, as they approach the Cauchy 
horizon.
If they develop an infinite blue-shift along the Cauchy horizon, this in
some sense is like saying this horizon would be `destroyed' and the
spacetime region beyond, which is exposed to the naked singularity,
would no longer be accessible. Hence predictability will be preserved in
the observable spacetime region. Interestingly enough, it has been found
that the Cauchy horizon in the dust and perfect fluid examples does not
have a blue-shift instability.

\noindent{\it Quantum effects}: The censorship hypothesis as such is
concerned with the nature of singularities in classical general relativity.
However, even if naked singularities do occur in the classical theory, 
one could ask if their formation would be avoided when quantum effects
in their vicinity are taken into account. This would be a quantum cosmic
censorship. Some investigations have taken place in this direction
[38, 39] and this very interesting question deserves to be pursued further.
Essentially the idea is to repeat for a naked singularity the kind of
calculation Hawking carried out to show that quantum effects cause 
black-holes to radiate. Since regions of very high curvature are exposed
near the naked singularity, intense particle production can be expected.
It is typically found in these calculations that as a result of the
produced particles, the energy-momentum flux at infinity diverges - in
spite of the fact that the naked singularity is massless and the classical
outgoing geodesics have infinite redshift! Although back-reaction
calculations are hard to carry out, the infinite flux would suggest that
the naked singularity formation will be avoided. From the quantum viewpoint,
naked singularities appear to be explosive events, and the outgoing flux 
might be their only possible observational signature. It is worth studying
the properties of this flux in detail to understand what observations, 
if any, can detect naked singularities if they do occur in nature. 
 
Thus we find that properties like curvature strength, masslessness,
redshift, blue-shift instability, and quantum effects give a mixed 
sort of picture regarding the significance of these naked singularity 
examples. An optimistic assessment of this situation is that there is a
good deal of richness in the problem, and much to think about, before
we can decide one way or the other.   

\vskip 0.2 in

\centerline{\bf 3. Non-spherical gravitational collapse}

\smallskip

\noindent As we have seen, there are examples of naked singularities
in spherical collapse. By assuming that the evolution can be continued
beyond the Cauchy horizon, one can conclude that the collapsing star will
eventually shrink below its Schwarzschild radius and the event horizon 
will form (according to the infalling observer).
There is also evidence that the horizon is stable to small perturbations.
However, if there are large departures from spherical symmetry, the
picture could be different, and the horizon may not form at all. The
naked singularity so forming would qualitatively be of a different kind,
compared to the ones seen in spherically symmetric spacetimes.

Our knowledge of exact solutions in the non-spherical case is inevitably
even more limited than for spherical systems, and one must again resort
to introducing some symmetry. An important early study was due to
Thorne [40], and was motivated by the work of Lin, Mestel and Shu [41]
on the collapse of Newtonian spheroids. Thorne examined the collapse of
an infinite cylinder and showed that it develops a curvature singularity
without an event horizon forming. Considerations such as these led him
to propose what came to be known as the hoop conjecture, which he stated
as follows [40]:

\begin{itemize}
\item Horizons form when and only when a mass $M$ gets compacted into
a region whose circumference in EVERY direction is 
{\cal C} \raisebox{-.6ex}{${\buildrel <    \over \sim}$} $4\pi M.$
\end{itemize}

\noindent According to the conjecture, collapsing objects which become
so asymmetric as to attain a circumference which is greater than the bound
will not develop horizons; hence if a singularity forms it will be naked.
We note that even if the conjecture holds, a naked singularity can form,
(as it does sometimes in spherical collapse), but an event horizon will
also form. One could say that the naked singularities which form when the
hoop conjecture holds are of a less serious nature than those which form
when the conjecture does not hold. In the former case an infalling observer
cannot communicate with asymptotic observers after crossing the horizon, 
while in the latter case no such restriction arises. 

Important numerical simulations were carried out by Shapiro and
Teukolsky [42] to test the hoop conjecture. They studied the
gravitational collapse of homogeneous non-rotating oblate and prolate 
spheroids of collisionless gas, starting from rest. Maximal slicing is 
used, and the 
evolution of the matter particles is followed with the help of the Vlasov 
equation in the self-consistent gravitational field. The development of a
singularity is detected by measuring the Riemann invariant at every point 
on the spatial grid. Since an event horizon can be observed only by
tracking null rays indefinitely, they instead search for the formation of
an apparent horizon (the boundary of trapped surfaces) - the
existence of the apparent horizon can be determined locally. If a spacetime
region has an apparent horizon, it will also have an event horizon, to its
outside.

They found that oblate spheroids first collapse to thin pancakes, but
then the particles overshoot and ultimately the distribution becomes
prolate and collapses to a thin spindle. An apparent horizon develops
to enclose the spindle, which eventually becomes singular, and a black
hole is formed. The minimum exterior circumference in the polar and
equatorial directions is consistent with the requirements of the hoop 
conjecture. The collapse of prolate 
spheroids leads however to the formation of a spindle singularity, with
no evidence for an apparent horizon covering the singularity. The initial
dimensions of the spheroid are such that the minimum circumference, at
the time of formation of the singularity, exceeds $4\pi M$. The collapse
of smaller prolate spheroids leads to spindle singularities that are
covered by a horizon, again favouring the conjecture.
Shapiro and Teukolsky suggested, noting the absence of an apparent
horizon for large prolate configurations, that the resulting spindle
singularity is naked, and hence that the hoop conjecture holds.

They were also careful to point out that the absence of an apparent
horizon up until the time of singularity formation does not necessarily
imply the absence of an event horizon. Hence, strictly one could not 
conclude that the singularity is necessarily naked. Wald and Iyer [43] 
showed this
mathematically with an example - they demonstrated that Schwarzschild
spacetime can be sliced with nonspherical slices, which approach 
arbitrarily close to the singularity, but do not have any trapped
surfaces. Another analytical example showing that absence of apparent
horizon does not imply the singularity is naked can be found in 
[36], where inhomogeneous spherical dust collapse was studied.
However, in support of their conclusion Shapiro and Teukolsky
pointed out that null rays continue to propagate away from the region of
the singularity until when the simulations are terminated, and that the
formation of an event horizon is unlikely. It is perhaps fair to conclude
that while their numerical simulations are of major importance
and their results suggestive, further investigations are necessary to
settle the issue. An analytical demonstration analogous to these simulations
was worked out in [44]. 

Another interesting analytical example is the
quasi-spherical dust solution due to Szekeres, which also admits naked
singularities [45], including those having strong curvature [46].

\vfil\eject

\centerline{\bf 4. Discussion}

\smallskip

\noindent We now attempt a critical comparison of the results reviewed
here, and discuss their implications for the censorship hypothesis.
(For other recent reviews of cosmic censorship see Clarke [4] and
Joshi [47]). Let us begin with a quick summary, even though it amounts 
to repetition.

Very massive stars are expected to end their gravitational collapse in
a singularity. There has been around an unproven conjecture that the 
singularity will be hidden behind an event horizon, and hence such stars
will become black-holes. If the conjecture is false, some stars can end
up as naked singularities - this will have major implications for
black-hole physics and astrophysics, and for classical general relativity.
Since a proof for the conjecture has not been forthcoming, relatively modest
attempts have been made to study specific examples of gravitational
collapse. These studies, which so far have been mostly for spherical
collapse, have thrown up some surprises. The collapse does not always end
in a black-hole; for some initial data it ends in a naked singularity,
and this is true for various forms of matter.

The spherical gravitational collapse of inhomogeneous dust leads to
weak naked singularities for generic initial data; the strong naked
singularities which do form for some data are non-generic. The naked 
singularity, irrespective of whether it is weak or strong, is massless.
The outgoing geodesics have an infinite redshift, and the Cauchy 
horizon does not have a blue-sheet instability. The collapse of a
self-similar perfect fluid also exhibits strong curvature naked
singularities for some initial data - these are again massless, have
infinitely redshifted outgoing geodesics, and the Cauchy horizon is
stable. Numerical studies of scalar field collapse suggest that the critical
solution is a naked singularity; elsewhere in the data space the collapse
ends either in dispersal or a singularity. It seems unclear as to
whether this singularity is definitely covered, or could be naked. For
collapse of a general form of matter, there is an existence proof that
for some of the data the singularity will be naked and strong - its
genericity is an open issue. There appear to be no conclusive studies
on non-spherical perturbations of these solutions, or on non-spherical
collapse as such - the simulations of Shapiro and Teukolsky are however
an important progress in this direction.

A broad conclusion is that at this early stage in collapse studies, one 
does not have enough evidence to take a grand decision about the validity of
the censorship hypothesis, one way or the other. Further, it does not appear
very useful right now to try and prove a specific version of the
hypothesis. Consider, just as an example, the following proposal:

\begin{itemize}
\item Gravitational collapse of {\it physically reasonable matter} 
starting from {\it generic initial data} leads to the
formation of a black-hole or a naked singularity. The naked singularity
is massless and gravitationally weak, the Cauchy horizon does not have
a blue-shift instability and the redshift along outgoing geodesics 
is infinite.
\end{itemize}
\noindent This does not appear to be a very interesting proposal to prove, 
given the number of properties attached to the naked singularity, nor is it
certain that it will survive further studies of collapse models.  
It is not even clear whether this proposal proves or disproves cosmic 
censorship!

What is noteworthy however is that naked singularities {\it do} occur in
dynamical collapse, side by side with black-hole solutions, so to say.
In order to address the censorship hypothesis, one
has to assess the significance of their properties, and also generalise  
the models studied so far. There are perhaps three important questions: 
(i) what is the role of the form of matter? (ii) are the
naked singularities genuine features of the spacetime geometry? (iii) do
they have observational effects? We respond to these questions briefly.

The form of matter in these models is dust, a perfect fluid, a scalar
field, or where an existence proof has been given, a general $T_{ik}$
obeying the weak energy condition. There are quite a few views on using
dust as a form of matter in these studies, which we try to enumerate.
Firstly, since dust collapse can give rise to singularities even in
Newtonian gravity or in special relativity, it is said that the (naked)
singularities being observed in general relativity have nothing to do
with gravitational collapse. This view appears acceptable for 
shell-crossing and the weak shell-focussing naked singularities. But
it is difficult to accept it for the strong naked singularities which
crush physical volumes to zero and hence ought to be a genuine general
relativistic feature. Secondly, the evolution of collisionless matter is
described by the Einstein-Vlasov equations; at any given point in space
the particles have a distribution of momenta. Dust is a very special case
of these equations, defined by the assumption that all particles have
exactly the same momentum. It has been shown [48] that the spherically
symmetric Einstein-Vlasov system has global solutions which do not contain
singularities, naked or otherwise, and hence censorship is preserved. 
However, we recall that dust collapse itself has a rich structure,
admitting both black-holes and naked singularities, a variety in trapped
surface dynamics, and of course includes the classic Oppenheimer-Snyder
model of black-hole formation. It would be a little surprising if these dust
features turn out to have no connection at all with more realistic collapse
models, which do have singularities. Thirdly, it has been 
suggested, though certainly not universally accepted, that during late 
stages of collapse matter will effectively behave like dust. In my view, a 
useful attitude towards the dust collapse results is to treat them as a
learning exercise and see if they will survive when more general forms of
matter are considered.

Quite naturally, perfect fluids with pressure get more serious consideration
than dust. Yet, the naked singularities found in their collapse can be 
objected to by saying that a fluid description is phenomenological and not
fundamental. This objection has been weakened by the results showing
naked singularities in scalar field collapse, and also by the existence
proofs for naked singularities with a general form of $T_{ik}$. 
Also, the assumption of self-similarity made in the explicit examples 
given by Ori and Piran [13] could be considered as restrictive. The existence
proofs by Dwivedi and Joshi relax this assumption. It appears safe to
conclude at this stage that the form of matter is not crucial in the
examples of naked singularities that have been found so far.

The second question deserves a more serious consideration, and holds the
key to the validity of the censorship hypothesis. Are these naked
singularities genuine features of the spacetime geometry? What is the
relative importance of the properties like masslessness of the naked 
singularity, curvature strength, redshift along outgoing geodesics, and 
instability of the Cauchy horizon? Do one or more of these features 
establish that the naked singularity is not genuine? There is a need to 
develop what one might call the theory of naked singularities, in order to
answer a difficult and important question of this sort. Stability against
non-spherical perturbations will also help decide the significance of these
examples. The singularity theorems showed that gravitational singularities
are not restricted to spherical symmetry; if this is any guide then one
would expect both the black-hole as well as naked singularity solutions to
persist when sphericity is relaxed. It is indeed difficult to visualize how
naked singularities will convert themselves to black-holes when asymmetry
is introduced.

The third question, as to whether naked singularities might have
observational effects, also deserves attention in view of the examples
now available. In spite of limited theoretical evidence for their
formation, black-holes have been successfully used to model many observed
astrophysical processes. One did not have to wait for the censorship
hypothesis to be proved before applications of black-holes could begin.
In a similar vein, one ought to give naked singularities a chance, so as
to examine if they will or will not have a significant flux emission, due
to quantum effects or otherwise.

While there is no clearly defined line of attack for further investigations
of the censorship hypothesis, two specific approaches appear to hold
promise. Firstly, the methods used by Belinskii, Lifshitz and Khalatnikov
to prove the generality of singularities [49] involve construction of
solutions of Einstein equations near spacetime singularities. Perhaps one
could study propagation of light rays using these solutions, to investigate
if the singularities could be naked. Secondly, major advances in numerical
relativity have made the subject ripe for studies on the censorship
hypothesis. For instance, it should be possible to check numerically if the
naked singularities of spherical inhomogeneous dust collapse persist if the 
collapse is non-spherical. Or, to check for naked singularities in spherical
perfect fluid collapse, when the self-similarity assumption is relaxed.

How does a very massive star evolve during the final stages of its
collapse? Does it choose to hide behind the event horizon to die a
silent death, or does it explode dramatically, exposing the singularity,
as if the Big Bang was being reenacted for our benefit? Recent developments
in our understanding of classical general relativity leave room for 
both possibilities. Further studies on gravitational collapse should prove
to be exciting, and it remains to be seen whether naked singularities
will come to play a role in physics and astrophysics.

\bigskip

\leftline{\bf Acknowledgements}
\smallskip
\noindent It is a pleasure to thank Patrick Brady, Fred Cooperstock, 
Tathagata Dasgupta,
I. H. Dwivedi, Sanjay Jhingan, Pankaj Joshi, Kerri Newman, Amos Ori, 
Ed Seidel, Peter Szekeres, C. S. Unnikrishnan, Cenalo Vaz, 
Shwetketu Virbhadra and Louis Witten for helpful discussions. I would
also like to thank the Institute of Mathematical Sciences, Madras for
hospitality.

\vskip 0.2 in

\centerline{\bf REFERENCES}
\smallskip
\begin{description}

\item[{[1]}] \hspace*{0.06 in}
 R. Penrose (1965) Phys. Rev. Lett. {\bf 14} 57; S. W. 
Hawking (1967) Proc. Roy. Soc. Lond. A {\bf 300} 187; S. W. Hawking and 
Penrose (1970) Proc. Roy. Soc. Lond. A {\bf 314} 529.  

\item[{[2]}]\hspace*{0.06in} R. Penrose (1969) Rivista del 
Nuovo Cimento {\bf 1} 252.

\item[{[3]}]\hspace*{0.06in} R. Penrose (1972) 
Nature {\bf 236} 377; (1974) Ann. 
New York Acad. Sci. {\bf 224} 125.

\item[{[4]}]\hspace*{0.06in} C. J. S. Clarke (1993) Class. 
Quantum Grav. {\bf 10} 1375. 

\item[{[5]}]\hspace*{0.06in} S. W. Hawking and G. F. R. Ellis 
(1973) The large-scale structure
of space-time, Section 4.3 (Cambridge University Press).

\item[{[6]}]\hspace*{0.06in} R. C. Tolman (1934) Proc. Nat. Acad. Sci. 
USA {\bf 20} 169; H. Bondi (1947) Mon. Not. Astron. Soc. {\bf 107} 410.

\item[{[7]}]\hspace*{0.06in} J. R. Oppenheimer and H. Snyder 
(1939) Phys. Rev. {\bf 56} 455.

\item[{[8]}]\hspace*{0.06in} P. Yodzis, H.-J Seifert and 
H. M\"{u}ller zum Hagen (1973) Commun. Math. Phys. {\bf 34} 135;
(1974) Commun. Math. Phys. {\bf 37} 29. 

\item[{[9]}]\hspace*{0.06in} D. M. Eardley and L. Smarr (1979) 
Phys. Rev. D {\bf 19}, 
2239; D. Christodoulou (1984) Commun. Math. Phys. {\bf 93} 
171; R. P. A. C. Newman, Class. Quantum Grav. {\bf 3}, 527 
(1986); B. Waugh and K. Lake (1988) Phys. Rev. D {\bf 38} 1315; V. Gorini, 
G. Grillo and M. Pelizza (1989) Phys. Lett. A {\bf 135} 
154; G. Grillo (1991) Class. Quantum Grav. {\bf 8} 
739; R. N. Henriksen and K. Patel (1991) Gen. Rel. Gravn. {\bf 23} 
527; I. H. Dwivedi and S. Dixit (1991) Prog. Theor. Phys. {\bf 85} 
433; I. H. Dwivedi and P. S. Joshi (1992) Class. Quantum Grav.
{\bf 9} L69; P. S. Joshi and I. H. Dwivedi (1993) Phys. Rev. {\bf D47} 
5357; P. S. Joshi and T. P. Singh (1995) Phys. Rev. {\bf D51} 
6778; T. P. Singh and P. S. Joshi (1996) Class. Quantum Grav. 
{\bf13} 559; C. S. Unnikrishnan (1996) Phys. Rev. {\bf D53} 
R580; H. M. Antia, Phys. Rev. (1996) {\bf D53} 3472; Sanjay 
Jhingan, P. S. Joshi and T. P. Singh (1996) 
gr-qc/9604046; T. P. Singh (1996) The central singularity in spherical
dust collapse - a review, in preparation.

\item[{[10]}] R. P. A. C. Newman, Ref. 9 above.

\item[{[11]}] C. W. Misner and D. H. Sharp (1964) Phys. Rev. {\bf 136} B571.  

\item[{[12]}] E. M. Lifshitz and I. M. Khalatnikov (1961) 
Soviet Physics JETP {\bf 12} 108; M. A. Podurets (1966)
Soviet Physics - Doklady {\bf 11} 275.

\item[{[13]}] A. Ori and T. Piran, Phys. Rev. D (1990) {\bf 42} 1068.

\item[{[14]}] P. S. Joshi and I. H. Dwivedi (1992) 
Commun. Math. Phys. {\bf 146} 333.

\item[{[15]}] P. S. Joshi and I. H. Dwivedi (1993) Lett. Math. Phys. 
{\bf 27} 235. 

\item[{[16]}] P. Szekeres and V. Iyer (1993) Phys. Rev. {\bf D47} 4362.

\item[{[17]}] P. C. Vaidya (1943) Current Science {\bf 13} 183; (1951)
Physical Review {\bf 83} 10; (1951) 
Proc. Indian Acad. Sci. {\bf A33} 264.

\item[{[18]}] W. A. Hiscock, L. G. Williams and D. M. Eardley (1982)
Phys. Rev. {\bf D26} 
751; Y. Kuroda (1984) Prog. Theor. 
Phys. {\bf 72} 63; A. Papapetrou (1985) in A Random Walk in General 
Relativity, Eds. N. Dadhich, J. K. Rao, J. V. Narlikar and
C. V. Vishveshwara (Wiley Eastern, New Delhi); G. P. Hollier (1986) 
Class. Quantum Grav. {\bf 3} L111; W. Israel (1986) Can. Jour. Phys. 
{\bf 64} 120; K. Rajagopal and K. Lake (1987) Phys. Rev. {\bf D35}
1531; I. H Dwivedi and P. S. Joshi (1989) Class. Quantum 
Grav. {\bf 6} 1599; (1991) Class. Quantum Grav. 
{\bf 8} 1339; P. S. Joshi and I. H. Dwivedi (1992) Gen. Rel. 
Gravn. {\bf24} 129; J. Lemos (1992) Phys. Rev. Lett.
{\bf 68} 1447.

\item[{[19]}] D. Christodoulou (1986) Commun. Math. Phys. 
{\bf 105} 337, {\bf 106} 587; (1987) {\bf 109} 591,
{\bf 109} 613; (1991) Commun. Pure Appl. Math. {\bf XLIV}
339; (1993) {\bf XLVI} 1131; 
(1994) Ann. Math. {\bf 140} 607. 

\item[{[20]}] M. W. Choptuik (1994) in Deterministic Chaos in General
Relativity, Eds. D. Hobill, A. Burd and A. Coley (Plenum, New York).

\item[{[21]}] M. W. Choptuik (1993) Phys. Rev. Lett. {\bf 70} 9.

\item[{[22]}] C. Gundlach, J. Pullin and R. Price (1994) 
Phys. Rev. {\bf D49} 890; D. Garfinkle (1995) Phys. Rev. {\bf D51} 
5558; R. S. Hamade and J. M. Stewart (1996) Class. and 
Quantum Grav. {\bf 13} 497.

\item[{[23]}] M. D. Roberts (1989) Gen. Rel. Gravn. {\bf 21} 907; V. Husain,
E. Martinez and D. Nunez (1994) 
Phys. Rev. {\bf D50} 3783; J. Traschen (1994) Phys. Rev. {\bf D50}
7144; Y. Oshiro, K. Nakamura and A. Tomimatsu (1994)
Prog. Theor. Phys. {\bf 91} 1265; P. R. Brady (1994) 
Class. Quantum Grav. {\bf 11} 1255; (1995) Phys. Rev. {\bf D51} 4168.

\item[{[24]}] C. Gundlach (1995) Phys. Rev. Lett. 
{\bf 75} 3214; (1996) gr-qc/9604019.

\item[{[25]}] A. M. Abrahams and C. R. Evans (1993) Phys. Rev. Lett.
{\bf 70 2980}. 

\item[{[26]}] C. R. Evans and J. S. Coleman (1994) Phys. Rev. Lett.
{\bf 72} 1782.

\item[{[27]}]  D. Maison (1994) gr-qc/9504008.

\item[{[28]}]  E. W. Hirschmann and D. M. Eardley (1995) Phys. Rev. {\bf D51}
4198; (1995) gr-qc/9506078; (1995) gr-qc/9511052; D. M. Eardley, 
E. W. Hirschmann and J. H. Horne (1995) gr-qc/9505041; R. S. Hamade, 
J. H. Horne and J. M. Stewart (1995) gr-qc/9511024.

\item[{[29]}]  I. H. Dwivedi and P. S. Joshi (1994) Commun. Math. Phys. 
{\bf 166} 117. 

\item[{[30]}]  K. Lake (1992) Phys. Rev. Lett. {\bf 68} 3129.

\item[{[31]}]  F. J. Tipler (1977) Phys. Lett. {\bf A64} 8; F. J. Tipler,
C. J. S. Clarke and G. F. R. Ellis (1980)
in General Relativity and Gravitation (Vol. 2), Ed. A. Held
(Plenum, New York).  

\item[{[32]}]  C. J. S. Clarke (1993) Analysis of spacetime singularities
(Cambridge University Press).

\item[{[33]}]  C. J. S. Clarke and A. Krolak (1986) J. Geo. Phys.
{\bf 2} 127.

\item[{[34]}]  R. P. A. C. Newman, Ref. [9] above.

\item[{[35]}]  V. Gorini et al.; G. Grillo; B. Waugh and K. Lake; I. H. 
Dwivedi and P. S. Joshi; T. P. Singh and P. S.
Joshi; Ref. [9] above.

\item[{[36]}]  Sanjay Jhingan, P. S. Joshi and T. P. Singh, Ref. [9] above.

\item[{[37]}]  A. G. Doroshkevich, Ya. B. Zeldovich and I. Novikov (1966)
Soviet Physics JETP {\bf 22} 122; V. de la Cruz, J. E. Chase
and W. Israel (1970) Phys. Rev. Lett. {\bf 24} 423; R. H. Price (1972) 
Phys. Rev. {\bf D5} 2419.   

\item[{[38]}]  L. H. Ford and L. Parker (1978) Phys. Rev. {\bf D17}
1485; W. A. Hiscock et al. Ref. 18 above.

\item[{[39]}]  C. Vaz and L. Witten (1994) Phys. Lett. {\bf B325} 27; (1995)
Class. Quantum Grav. {\bf 12} 2607; (1995) gr-qc/9511018; 
(1996) gr-qc/9604064.

\item[{[40]}]  K. S. Thorne (1972) in Magic Without Magic: {\it John
Archibald Wheeler}, Ed. John R. Klauder, 
(W. H. Freeman and Co., San Francisco).

\item[{[41]}]  C. C. Lin, L. Mestel and F. H. Shu (1965) 
Ap. J. {\bf 142} 1431. 

\item[{[42]}]  S. L. Shapiro and S. A. Teukolsky (1991) 
Phys. Rev. Lett. {\bf 66} 994; Phys. Rev. (1992) 
{\bf D45} 2006.

\item[{[43]}]  R. M. Wald and V. Iyer (1991) Phys. Rev. {\bf D44} 3719.

\item[{[44]}]  C. Barrab\`{e}s, W. Israel and P. S. Letelier (1991)
Phys. Lett. {\bf A160} 41.

\item[{[45]}]  P. Szekeres (1975) Phys. Rev. {\bf D12} 2941.  

\item[{[46]}]  P. S. Joshi and A. Krolak (1996) gr-qc/9605033.

\item[{[47]}]  P. S. Joshi (1993) Global Aspects in Gravitation and 
               Cosmology, Clarendon Press, OUP, Oxford,
               Chapters 6 and 7.

\item[{[48]}]  A. D. Rendall (1992) Class. Quantum Grav. {\bf 9} L99; (1992)
in Approaches to Numerical Relativity
Ed. R. d'Inverno (Cambridge University Press); G. Rein and 
A. D. Rendall (1992) Commun. Math. Phys. {\bf 150} 561; G. Rein, 
A. D. Rendall and J. Schaeffer (1995) Commun. Math. Phys. {\bf 168} 467.  

\item[{[49]}]  V. A. Belinskii, J. M. Khalatnikov and E. M. Lifshitz,
(1970) Advances in Physics {\bf 19} 525.    

\end{description}

\end{document}